\documentstyle[11pt,amssymb,epsf]{article}

\def\ben{\begin{equation}}
\def\een{\end{equation}}

\let\a=\alpha    
    
  \let\n=\nu

\let\C=\Chi

\def\nn{\nonumber} \def\bd{\begin{document}} \def\ed{\end{document}}
\def\ds{\documentstyle} \let\fr=\frac \let\bl=\bigl \let\br=\bigr
\let\Br=\Bigr \let\Bl=\Bigl
\let\bm=\bibitem
\let\na=\nabla
\let\pa=\partial \let\ov=\overline
\newcommand{\be}{\begin{equation}}
\newcommand{\ee}{\end{equation}}
\def\ba{\begin{array}}
\def\ea{\end{array}}
\def\ft#1#2{{\textstyle{{\scriptstyle #1}\over {\scriptstyle #2}}}}
\def\fft#1#2{{#1 \over #2}}
\def\del{\partial}
\def\vp{\varphi}
\def\sst#1{{\scriptscriptstyle #1}}
\def\oneone{\rlap 1\mkern4mu{\rm l}}
\def\td{\tilde}
\def\wtd{\widetilde}
\def\ie{\rm i.e.\ }
\def\dalemb#1#2{{\vbox{\hrule height .#2pt
        \hbox{\vrule width.#2pt height#1pt \kern#1pt
                \vrule width.#2pt}
        \hrule height.#2pt}}}
\def\square{\mathord{\dalemb{6.8}{7}\hbox{\hskip1pt}}}
\newcommand{\ho}[1]{$\, ^{#1}$}
\newcommand{\hoch}[1]{$\, ^{#1}$}
\newcommand{\bea}{\begin{eqnarray}}
\newcommand{\eea}{\end{eqnarray}}
\newcommand{\ra}{\rightarrow}
\newcommand{\lra}{\longrightarrow}
\newcommand{\Lra}{\Leftrightarrow}
\newcommand{\ap}{\alpha^\prime}
\newcommand{\bp}{\tilde \beta^\prime}
\newcommand{\tr}{{\rm tr} }
\newcommand{\Tr}{{\rm Tr} }
\def\0{{\sst{(0)}}}
\def\1{{\sst{(1)}}}
\def\2{{\sst{(2)}}}
\def\3{{\sst{(3)}}}
\def\4{{\sst{(4)}}}
\def\5{{\sst{(5)}}}
\def\6{{\sst{(6)}}}
\def\7{{\sst{(7)}}}
\def\8{{\sst{(8)}}}
\def\n{{\sst{(n)}}}
\def\cA{{{\cal A}}}
\def\cF{{{\cal F}}}
\def\tV{\widetilde V}
\def\tW{\widetilde W}
\def\tH{\widetilde H}
\def\tE{\widetilde E}
\def\tF{\widetilde F}
\def\tA{\widetilde A}
\def\im{{{\rm i}}}
\def\tY{{{\wtd Y}}}
\def\ep{{\epsilon}}
\def\vep{{\varepsilon}}
\def\R{\rlap{\rm I}\mkern3mu{\rm R}}
\def\bD{{{\bar D}}}

\def\R{\rlap{\rm I}\mkern3mu{\rm R}}
\def\bD{{{\bar D}}}
\def\R{{{\Bbb R}}}
\def\C{{{\Bbb C}}}
\def\H{{{\Bbb H}}}
\def\CP{{{\Bbb C}{\Bbb P}}}
\def\RP{{{\Bbb R}{\Bbb P}}}
\def\Z{{{\Bbb Z}}}

\def\bA{{{\Bbb A}}}
\def\bB{{{\Bbb B}}}
\def\bC{{{\Bbb C}}}
\def\bD{{{\Bbb D}}}
\def\bE{{{\Bbb E}}}
\def\bR{{{\Bbb R}}}  
\def\bZ{{{\Bbb Z}}}
\def\Re{{{\frak{Re}}}}
\def\Im{{{\frak{Im}}}}
\def\cosec{{\,\hbox{cosec}\,}}
\def\Gm{{\Gamma_{\!\! -}}}
\def\Gp{{\Gamma_{\!\! +}}}
\def\stan{{standard }}
\def\nonstan{{supernumerary }}
\def\btheta{{\bar\theta}}

\newcommand{\tamphys}{\it Center for Theoretical Physics,
Texas A\&M University, College Station, TX 77843, USA}
\newcommand{\umich}{\it Michigan Center for Theoretical Physics,
University of Michigan\\ Ann Arbor, MI 48109, USA}
\newcommand{\upenn}{\it Department of Physics and Astronomy,
University of Pennsylvania\\ Philadelphia,  PA 19104, USA}
\newcommand{\SISSA}{\it  SISSA-ISAS and INFN, Sezione di Trieste\\
Via Beirut 2-4, I-34013, Trieste, Italy}

\newcommand{\ihp}{\it Institut Henri Poincar\'e\\
  11 rue Pierre et Marie Curie, F 75231 Paris Cedex 05}

\newcommand{\damtp}{\it DAMTP, Centre for Mathematical Sciences,
 Cambridge University\\ Wilberforce Road, Cambridge CB3 OWA, UK}
\newcommand{\itp}{\it Institute for Theoretical Physics, University of
California\\ Santa Barbara, CA 93106, USA}

\newcommand{\auth}{Z.W. Chong\hoch{\ddagger}, 
M. Cveti\v{c}\hoch{\dagger}, G.W. Gibbons\hoch{\sharp},
H. L\"u\hoch{\star}, C.N. Pope\hoch{\ddagger} and 
   P. Wagner\hoch{\ddagger}}

\begin{document}

\begin{flushright}
\hfill{DAMTP-2001-25}\ \ \ {CTP TAMU-06/02}\ \ \ {UPR-986-T}\ \ \
{MCTP-02-20}\\
{March 2002}\ \ \
{hep-th/0204064}
\end{flushright}


\begin{center}
{ \large {\Large\bf General Metrics of $G_2$ Holonomy and Contraction
Limits}}

\vspace{20pt}
\auth

\vspace{3pt}
{\hoch{\dagger}\upenn}

\vspace{3pt}


\vspace{3pt}
{\hoch{\sharp}\damtp}

\vspace{3pt}
{\hoch{\star}\umich}

\vspace{3pt}
{\hoch{\ddagger}\tamphys}

\vspace{3pt}

\underline{ABSTRACT}
\end{center}

    We obtain first-order equations for $G_2$ holonomy of a wide class
of metrics with $S^3\times S^3$ principal orbits and $SU(2)\times
SU(2)$ isometry, using a method recently introduced by Hitchin.  The
new construction extends previous results, and encompasses all
previously-obtained first-order systems for such metrics.  We also
study various group contractions of the principal orbits, focusing on
cases where one of the $S^3$ factors is subjected to an Abelian,
Heisenberg or Euclidean-group contraction.  In the Abelian
contraction, we recover some recently-constructed $G_2$ metrics with
$S^3\times T^3$ principal orbits.  We obtain explicit solutions of
these contracted equations in cases where there is an additional
$U(1)$ isometry.  We also demonstrate that the only solutions of the
full system with $S^3\times T^3$ principal orbits that are complete
and non-singular are either flat $\R^4$ times $T^3$, or else the
direct product of Eguchi-Hanson and $T^3$, which is asymptotic to
$\R^4/\bZ_2\times T^3$.  These examples are in accord with a general
discussion of isometric fibrations by tori which, as we show, in
general split off as direct products.  We also give some (incomplete)
examples of fibrations of $G_2$ manifolds by associative 3-tori with
either $T^4$ or K3 as base.

{\vfill\leftline{}\vfill
\vskip 5pt
\footnoterule
{\footnotesize \hoch{\dagger} Research supported in part by DOE grant
DE-FG02-95ER40893 and NATO grant 976951. \vskip -12pt} \vskip 14pt
{\footnotesize \hoch{\star} Research supported in full by DOE grant
DE-FG02-95ER40899 \vskip -12pt} \vskip 14pt
{\footnotesize  \hoch{\ddagger} Research supported in part by DOE
grant DE-FG03-95ER40917.\vskip  -12pt}}

\pagebreak
\setcounter{page}{1}

\tableofcontents
\addtocontents{toc}{\protect\setcounter{tocdepth}{3}}
\vfill\eject

\section{Introduction}

   Manifolds $M_7$ of $G_2$ holonomy are of considerable interest
because they allow one to construct supersymmetric backgrounds of the
form (Minkowski)$_4\times M_7$ in M-theory.  Explicit examples of
complete, regular, non-compact $G_2$ metrics exist.  All the known
examples are of cohomogeneity one, with principal orbits that are
$S^3\times S^3$, $\CP^3$ or $SU(3)/(U(1)\times U(1))$.  The first
regular examples were found in \cite{brysal,gibpagpop}; these are
asymptotically conical (AC).  The most interesting case for physical
purposes is when the principal orbits are $S^3\times
S^3$.\footnote{The $G_2$ metrics with the other two types of principal
orbits can be viewed as Gromov-Hausdorff limits of a new class of ALC
Spin(7) metrics constructed in \cite{cglpspin7}.} More general systems
of equations for such metrics of $G_2$ holonomy were obtained in
\cite{cglp6fun,bragugogo}, and an explicit new solution, which is
asymptotically locally conical (ALC), was given in \cite{bragugogo}.
A rather general system of first-order equations for such metrics of
$G_2$ holonomy was obtained in \cite{brand,munify}.  Although the
general solution was not found in \cite{brand,munify}, it was shown
that three types of regular metrics could arise, in which the orbits
degenerate to $S^2$ \cite{cglpmcon}, $S^3$
\cite{bragugogo,brand,munify} or $T^{1,1}$ \cite{munify} at short
distance.  Classes of such metrics that are asymptotically locally
conical were found; these were denoted as $\bD_7$, $\bB_7$ and
$\wtd\bC_7$ respectively.  They all have a non-trivial parameter (two
for $\wtd\bC_7$) that adjusts the radius of the asymptotic circle
relative to the overall scale-size of the metric.  The
Gromov-Hausdorff limits of $\bD_7$ and $\bB_7$ are the resolved and
the deformed conifolds respectively, whilst the Gromov-Hausdorff
limits of $\wtd \bC_7$ give the family of Ricci-flat K\"ahler metrics
on the complex line bundle over $S^2\times S^2$ \cite{munify}.

   In a recent paper, Hitchin has given a new construction of certain
types of metrics of special holonomy, including seven-dimensional
metrics of $G_2$ holonomy \cite{hitch}.  The procedure involves
constructing diffeomorphism-invariant functionals on certain
differential forms.  By extremising the functionals, he obtains
Hamiltonian flow equations that lead to metrics of $G_2$ holonomy.  In
\cite{hitch}, as an application of the method, a new derivation of a
class of $G_2$ metrics previously obtained in
\cite{cglp6fun,bragugogo} was given.

   In section 2 of this paper, we apply Hitchin's procedure with a
somewhat more general starting point, and thereby we obtain a system
of first-order equations for metrics of $G_2$ holonomy and $S^3\times
S^3$ principal orbits that is more general than any obtained
hitherto.\footnote{The same generalisation of Hitchin's 3-form was
considered also in \cite{brand}, where the conditions for $G_2$
holonomy were expressed in the form of some coupled second-order
equations that are presumably equivalent to our first-order system.}
As we show in section 3, it not only encompasses the system of
first-order equations obtained in \cite{cglp6fun,bragugogo}, but also
the inequivalent system obtained in \cite{brand,munify}.
Additionally, it encompasses a recently-obtained class of $G_2$
metrics with $S^3\times T^3$ principal orbits \cite{guyaza}.  We shall
show how this first-order system arises as a contraction limit of our
new results for $S^3\times S^3$ orbits, in which the abelian limit of
$S^3$ is taken.

   We then turn in section 4 to a more general consideration of
contraction limits.  Specifically, we can apply various group
contractions to the $S^3=SU(2)$ factors.  Some earlier discussion of
this procedure in relation to metrics of special holonomy was given in
\cite{gilupost}.

   In section 5 we study the solutions of the first-order equations
that arise in the contraction limits.  These equations are simpler
than those associated with the uncontracted $S^3\times S^3$ orbits,
and so they are typically more tractable from the viewpoint of
obtaining exact and explicit solutions.  In particular, we study the
solutions of the metrics with $S^3\times T^3$ principal orbits 
in cases where there is an extra $U(1)$ isometry.  We explicitly
show that the metrics are all singular, except for flat $\R^4$ times
$T^3$, or else Eguchi-Hanson times $T^3$.  The latter can be viewed
as a Gromov-Hausdorff limit of the generic singular metrics, in 
which the radius of one of the circles in $T^3$ goes to zero.  
The manifold in this case has the form $\R^4/\bZ_2 \times T^3$ 
at large distance.

   In section 6, we discuss the solutions of the more general
triaxial case with $S^3 \times T^3$ orbits and no $U(1)$ isometry, 
and we show that again flat $\R^4$ times $T^3$, and Eguchi-Hanson 
times $T^3$, are the only complete and non-singular metrics. 

   In section 7 we give more general arguments to show that in 
isometric fibrations by tori, the toroidal directions always, in
general, split off as a direct-product factor.

   In section 8, we give some (incomplete) examples of fibrations of
$G_2$ manifolds by associative 3-tori, with either $T^4$ or K3 as
base.

   Finally, we remark that since $G_2$ and Spin(7) manifolds provide
natural compactifications in M-theory \cite{acharaya,atmava,atiwit},
there has been a considerable effort recently in constructing explicit
non-compact $G_2$ and Spin(7) metrics (see also the additional
references \cite{kanyas1}-\cite{hersfe}).

\section{New $G_2$ metrics with $S^3\times S^3$ principal orbits}
\label{newg2}

   To obtain the new $G_2$ metrics we generalise the example
considered by Hitchin in \cite{hitch}.  We take the 3-form and 4-form
used in the general construction in \cite{hitch} to be given by
\bea
\rho &=& n\, \Sigma_1\, \Sigma_2\, \Sigma_3 - m\, \sigma_1\, \sigma_2\,
\sigma_3 + x_1\, (\sigma_1\, \Sigma_2\, \Sigma_3 - \sigma_2\, \sigma_3
\, \Sigma_1) + \hbox{2 cyclic terms}\,,\nn\\
\sigma &=&y_1\, \sigma_2\, \Sigma_2 \, \sigma_3\, \Sigma_3 +
y_2\, \sigma_3\, \Sigma_3 \, \sigma_1\, \Sigma_1 +
y_3\, \sigma_1\, \Sigma_1 \, \sigma_2\, \Sigma_2\,,\label{rhosig}
\eea
where $\Sigma_i$ and $\sigma_i$ are two sets of left-invariant 1-forms
of $SU(2)$.  (The example considered in \cite{hitch} had $m=n=1$, and
as we shall see, this choice significantly restricts the generality of
the results.)  From (\ref{rhosig}), the next step is to calculate the
associated potentials $V(\rho)$ and $W(\sigma)$, whose general
definitions were given in \cite{hitch}.

    Since previous discussions of the approach in \cite{hitch} have
been somewhat abstract, it is perhaps worthwhile to present the key
calculational steps here ``with indices.''  To obtain $V(\rho)$, we
first define
\be
K_a{}^b\equiv \ft1{12} \rho_{c_1 c_2 c_3}\, \rho_{c_4 c_5 a}\, 
\vep^{c_1 c_2 c_3 c_4 c_5 b}\,,
\ee
where $\vep^{c_1\cdots c_6}$ is the Levi-Civita 
tensor density in 6-dimensions with values $\pm1$ and 0.  Then,
$V(\rho)$ is given by
\be
V(\rho) = \sqrt{-\ft16 K_a{}^b\, K_b{}_{\phantom{\Sigma}}^a}\,.
\ee
To calculate $W(\sigma)$, we first construct the dual tensor density
\be
\td \sigma^{ab} \equiv \ft1{24}\, \vep^{ab c_1c_2c_3c_4}\, 
\sigma_{c_1c_2c_3c_4}\,.
\ee
From this, $W(\sigma)$ is calculated from
\be
W(\sigma)^2 = \ft1{48}\, \vep_{c_1\cdots c_6}\, \td\sigma^{c_1 c_2}\,
\td\sigma^{c_3c_4}\,\td\sigma^{c_5c_6}\,.
\ee

   One now defines the Hamiltonian $H=V(\rho) - 2 W(\sigma)$.  It is
shown in \cite{hitch} that a metric of $G_2$ holonomy is obtained if
the first-order equations following from the Hamiltonian flow
\be
\dot x_i = -\fft{\del H}{\del y_i}\,,\qquad \dot y_i = 
     \fft{\del H}{\del x_i}\,,\label{hamflow}
\ee
are satisfied,\footnote{As in \cite{hitch}, there is a natural pairing
between the invariant 3-form and 4-form that is non-degenerate, and
the symplectic form is just a multiple of $dx_i\wedge dy_i$.} where
the dot denotes a derivative with respect to an additional ``time''
variable $t$.  The metric itself is obtained as follows.  First, one
takes the ``square root'' of the 4-form $\sigma$, writing it as
$\sigma=\ft12 \omega^2$. Then, the associative 3-form of the $G_2$
metric is given by \footnote{This associative 3-form was also
considered in \cite{gukov}.}
\be
\Phi_\3= dt\wedge \omega + \rho\,.\label{phidef}
\ee
From this, one calculates the $G_2$ metric as follows.  First, we
define the symmetric tensor density
\be 
B_{AB} = -\ft1{144} \Phi_{A C_1 C_2}\, \Phi_{B C_3 C_4}\,
\Phi_{C_5 C_6 C_7}\, \vep^{C_1\cdots C_7}\,,\label{dendef} 
\ee
where $\vep^{C_1\cdots C_7}$ is the Levi-Civita tensor density in
seven dimensions.  The metric tensor is then given by
\be
g_{AB} = \det(B)^{-1/9}\, B_{AB}\,.\label{metdef}
\ee

   Applying this construction to (\ref{rhosig}), we find that the
2-form $\omega$ can be taken to be
\be
\omega = \sqrt{\fft{y_2\, y_3}{y_1}}\, \sigma_1\wedge \Sigma_1
+ \sqrt{\fft{y_3\, y_1}{y_2}}\, \sigma_2\wedge \Sigma_2 +
\sqrt{\fft{y_1\, y_1}{y_3}}\, \sigma_3\wedge \Sigma_3\,,
\ee
and it is easily verified that this satisfies the criterion 
$\omega\wedge\rho=0$ that is necessary for the 
$SL(3,\bC)$ reduction by $\rho$ and the $Sp(6,\bR)$ reduction 
by $\sigma$ to intersect in $SU(3)$ \cite{hitch}. Writing
$V(\rho)=\sqrt{-U}$, we find that the potentials are given by
\bea
U&=& m^2\, n^2 -2m\, n\, (x_1^2+x_2^2+x_3^2) - 4(m+n)\,
x_1\, x_2\, x_3\nn\\
&&+ x_1^4+x_2^4+x_3^4 - 2x_1^2\, x_2^2 - 2 x_2^2\, x_3^2 -2 x_3^2\, 
x_1^2\,,\label{udef}\\
W(\sigma) &=& \sqrt{y_1\, y_2\, y_3}\,.
\eea
Thus the first-order equations following from (\ref{hamflow}) are
\be
\dot x_1 = \sqrt{\fft{y_2\, y_3}{y_1}}\,,\qquad
\dot y_1 = \fft{m\, n\, x_1 + (m+n)\, x_2\, x_3 
+ x_1\, (x_2^2+x_3^2-x_1^2)}{\sqrt{y_1\, y_2\, y_3}}\,,\label{newfo}
\ee
and cyclically for the 2 and 3 directions.  We have used the
Hamiltonian constraint $H=0$, \ie 
\be
U=-4 y_1\, y_2\, y_3\,,\label{hamilton}
\ee
in writing the $\dot y_i$ equations.  Using (\ref{metdef}), we find
that the metric is given by
\bea
ds^2 &=& dt^2 + \fft1{y_1}\, \Big[ (n\, x_1 + x_2\, x_3)\, \Sigma_1^2 +
 (m\, n + x_1^2 - x_2^2 - x_3^2)\, \Sigma_1\, \sigma_1 +
   (m\, x_1 + x_2\, x_3)\, \sigma_1^2\Big]\nn\\
&&+\fft1{y_2}\, \Big[ (n\, x_2 + x_3\, x_1)\, \Sigma_2^2 +
 (m\, n + x_2^2 - x_3^2 - x_1^2)\, \Sigma_2\, \sigma_2 +
   (m\, x_2 + x_3\, x_1)\, \sigma_2^2\Big]\nn\\
&&+\fft1{y_3}\, \Big[ (n\, x_3 + x_1\, x_2)\, \Sigma_3^2 +
 (m\, n + x_3^2 - x_1^2 - x_2^2)\, \Sigma_3\, \sigma_3 +
   (m\, x_3 + x_1\, x_2)\, \sigma_3^2\Big]\,,\label{newmet}
\eea

   As an alternative demonstration that the first-order equations
(\ref{newfo}) do indeed imply $G_2$ holonomy, we can simply verify
that $d\Phi_\3=0$ and $d{*\Phi_\3}=0$, where $\Phi_\3$ is given by
(\ref{phidef}) and the Hodge dual is evaluated using the metric
(\ref{newmet}), which is derived from (\ref{dendef}) and
(\ref{metdef}).  After a mechanical calculation, one finds that the
dual 4-form $\Psi_\4\equiv {*\Phi_\3}$ is given by
\bea 
\Psi_\4&=& \Psi_{0123}\, dt\, \Sigma_1\, \Sigma_2\, \Sigma_3 +
\Psi_{0456}\, dt\, \sigma_1\, \sigma_2\, \sigma_3\nn\\
&& + \Psi_{0156}\, dt\,\Sigma_1\,\sigma_2\,\sigma_3 +
\Psi_{0234}\, dt\,\Sigma_2\, \Sigma_3\, \sigma_1 + \Psi_{2356} \,
\Sigma_2\, \Sigma_3\, \sigma_2\, \sigma_3 +\hbox{cyclic}\,,
\eea
where two further sets of 3 terms are added in the second line, 
cycled on (1,2,3) and (4,5,6) simultaneously.  The non-zero components
of $\Psi$ are given by
\bea
&&\Psi_{0123}=\fft1{\sqrt{-U}}\, [m\, n^2 - n\, (x_1^2+x_2^2+x_3^2) -
2 x_1\, x_2\, x_3]\,,\nn\\
&&\Psi_{0456}=\fft1{\sqrt{-U}}\, [m^2\, n - m\, (x_1^2+x_2^2+x_3^2) -
2 x_1\, x_2\, x_3]\,,\nn\\
&& \Psi_{0156}= \fft{1}{\sqrt{-U}}\, [m\, n \, x_1 +2m \, x_2\, x_3 
    +x_1\, (x_2^2 + x_3^2 -x_1^2)]\,,\nn\\
&&\Psi_{0234}= \fft{1}{\sqrt{-U}}\, [m\, n \, x_1 +2n \, x_2\, x_3 
    +x_1\, (x_2^2 + x_3^2 -x_1^2)]\,,\nn\\
&&
\Psi_{2356} = -\ft12 \sqrt{-U}\, \sqrt{\fft{y_1}{y_2\, y_3}}\,,
\eea
together with those following from simultaneous 
cycling on (1,2,3) and (4,5,6), with $U$ defined by (\ref{udef}).

    It is clear that $d\Phi_\3=0$ immediately implies the equations for
$\dot x_i$ in (\ref{newfo}).  From $d\Psi_\4=0$ we obtain
\be
\fft{d\Psi_{2356}}{dt} + \Psi_{0156} + \Psi_{0234}=0\,,
\ee
and cyclic permutations.  Taking linear combinations, and using the
$\dot x_i$ equations, these imply
\be
\dot y_1 = \fft{\sqrt{y_1\, y_2\, y_3}}{U} \, [m\, n\, x_1 + 
(m+n)\, x_2\, x_3 + x_1\, (x_2^2+x_3^2-x_1^2)]\,,\label{yeqs}
\ee
and cyclic permutations.  From these and the $\dot x_i$ equations we
can then establish that $(y_1\, y_2\, y_3)/U$ is a constant, which
without loss of generality may be chosen by scaling to be $-\ft14$, as
in (\ref{hamilton}).  Using this, (\ref{yeqs}) can be reduced to the
previous expressions given in (\ref{newfo}).  Thus we have confirmed,
by directly requiring the closure and co-closure of $\Phi_\3$, that
the metrics (\ref{newmet}) indeed have $G_2$ holonomy if the
first-order equations (\ref{newfo}) are satisfied.

\section{Specialisations to previous results}

    In this section, we show how the $G_2$ metrics obtained above 
reduce to various previously-known cases.

\subsection{Reduction to triaxial 6-function metrics}\label{sixfnsec}

   In \cite{cglp6fun,bragugogo}, a class of metrics given by
\be
ds^2 = dt^2 + a_i^2 (\Sigma_i-\sigma_i)^2 + b_i^2\, 
(\Sigma_i+\sigma_i)^2
\label{sixfn}
\ee
was considered.  It was shown that the metrics have $G_2$ holonomy if
the six functions $a_i$ and $b_i$ satisfy the first-order equations
\bea
\dot a_1 &=& \fft{a_1^2}{4 a_3\, b_2} + \fft{a_1^2}{4 a_2\, b_3} 
-\fft{a_2}{4b_3} - \fft{a_3}{4b_2} - \fft{b_2}{4 a_3} -\fft{b_3}{4a_2}
\,,\nn\\
\dot b_1 &=& \fft{b_1^2}{4 a_2\, a_3} - \fft{b_1^2}{4 b_2\, b_3} 
-\fft{a_2}{4a_3} - \fft{a_3}{4a_2} + \fft{b_2}{4 b_3} +\fft{b_3}{4b_2}
\,,\label{6funeq}
\eea
and cyclically in $(1,2,3)$.  These metrics have $SU(2)\times SU(2)$ 
isometry. 

   Comparing with our results in section \ref{newg2}, we can see that
the triaxial six-function metrics arise from the specialisation in
which we set $m=n$.  This is the case that was analysed in
\cite{hitch}.  The functions $x_i$ and $y_i$ in section \ref{newg2} 
are then given by
\bea
&&x_1= a_1\, a_2\, a_3 + a_3\, b_1\, b_2 + a_2\, b_1\, b_3 -
a_1\, b_2\, b_3\,,\nn\\
&&x_2= a_1\, a_2\, a_3 + a_3\, b_1\, b_2 - a_2\, b_1\, b_3 +
a_1\, b_2\, b_3\,,\nn\\
&&x_3= a_1\, a_2\, a_3 - a_3\, b_1\, b_2 + a_2\, b_1\, b_3 +
a_1\, b_2\, b_3\,,\nn\\
&&y_1= 4 a_2\, a_3\, b_2\, b_3\,,\quad
y_2= 4 a_1\, a_3\, b_1\, b_3\,,\quad
y_3=4 a_1\, a_2\, b_1\, b_2\,,
\eea
and $m$ and $n$, which are equal, are related to the $a_i$ and $b_i$ 
by 
\be
m=n =-a_1\, a_2\, a_3 + a_3\, b_1\, b_2 + a_2\, b_1\, b_3 +
a_1\, b_2\, b_3\,.\label{mcon1}
\ee
It can be verified also that the Hamiltonian constraint
(\ref{hamilton}) is identically satisfied.  Note that after setting
$m=n$ the expression (\ref{udef}) for $U$ factorises, to give
\be
U=(m-x_1-x_2-x_3)\, (m+x_1+x_2-x_3)\, (m+x_1-x_2+x_3)\, 
  (m-x_1 +x_2 + x_3)\,.
\ee
(In this special case with $m=n$, the actual value of $m$ is a trivial
overall scale parameter, which can be set to $m=1$ as in
\cite{hitch}.)  The equation (\ref{mcon1}) can be understood directly
from the six-function equations (\ref{6funeq}), which imply that the
cubic function in (\ref{mcon1}) is a constant of the motion.

       The associative 3-form now can be expressed as
\be
\Phi_\3 =e^{014} + e^{025} + e^{036} - e^{123} + e^{156} -
e^{246} + e^{345}\,,\label{special3form}
\ee
where $e^{ijk}\equiv e^i\wedge e^j\wedge e^k$, and the vielbein is
defined by
\be
e^0=dt\,,\qquad e^i=a_i\, (\Sigma_i-\sigma_i)\,,\qquad
e^{i+3} = b_i\, (\Sigma_i + \sigma_i)\,,\qquad
i=1,2,3\,.
\ee

   A four-function specialisation of (\ref{sixfn}), in which $a_1=a_2$
and $b_1=b_2$, includes in its solutions a family of complete and
regular ALC metrics with a minimal $S^3$.  In the Gromov-Hausdorff
limit this approaches the product of a circle and the deformed
conifold.

\subsection{Reduction to the conifold-unification metrics}
\label{conifoldsec}

   In \cite{cglpmcon} a new class of $G_2$ metrics with $S^3\times
S^3$ principal orbits was obtained, which includes regular solutions
that describe the resolved conifold in the Gromov-Hausdorff limit.
The class was extended further in \cite{brand,munify}, to a system
that encompasses \cite{cglpmcon} and also the previous four-function
specialisation described in section \ref{sixfnsec}.  In particular,
the system found in \cite{brand,munify} provides a unification, via
M-theory, of the deformed and resolved conifolds \cite{munify}.

   For the present purposes it is best to use the metric 
parameterisation in eqn (15) of \cite{munify} (with the tildes 
omitted):
\bea
ds_7^2 &=& dt^2 +  a^2\, [(\Sigma_1 +  g\, \sigma_1)^2 +
                          (\Sigma_2 +  g\, \sigma_2)^2]
+  b^2\,  [(\Sigma_1 -  g\, \sigma_1)^2 +
                          (\Sigma_2 -  g\, \sigma_2)^2]\nn\\
&&+  c^2\, (\Sigma_3-\sigma_3)^2 +  f^2\, (\Sigma_3 +  g_3\,
\sigma_3)^2\,.\label{d7ans2}
\eea
With
respect to the vielbein basis
\bea
&& e^0=dt\,,\quad  e^1= a\, (\Sigma_1 +  g\,
\sigma_1)\,,\quad  e^2= a\, (\Sigma_2 +  g\,
\sigma_2)\,,\quad  e^3 = c\, (\Sigma_3 - \sigma_3)\,,\nn\\
&& e^4 =  b\, (\Sigma_1 -  g\, \sigma_1)\,,\quad
 e^5 =  b\, (\Sigma_2 -  g\, \sigma_2)\,,\quad
 e^6 =  f\, (\Sigma_3 +  g_3\, \sigma_3)\,,\label{viel2}
\eea
the associative 3-form takes the same form as (\ref{special3form}).
The conditions for $G_2$ holonomy, $d\Phi_\3=0$ and $d{*\Phi_\3}=0$,
then imply the algebraic relation
\be
 g_3 =  g^2 - \fft{ c\, ( a^2- b^2)
                 (1- g^2)}{2 a\,  b\,  f}\,,
\label{constr2}
\ee
together with the first-order equations 
\bea
\dot a &=& \fft{ c^2\, ( a^2 - b^2) +
[4 a^2\, ( a^2- b^2)-  c^2\, (5  a^2- b^2) - 4
a\,  b\,  c\,  f]\,  g^2}{16 a^2\,  b\,  c\, 
g^2}\,,\nn\\
\dot b &=& -\, \fft{ c^2\, ( a^2- b^2) + [4  b^2\,
( a^2 - b^2) + c^2\, (5 b^2 -  a^2) - 4 a\,  b\,
 c\,  f]\, g^2}{16  a\,  b^2\,  c\,  g^2}\,,\nn\\
\dot c &=& \fft{ c^2 + ( c^2 -2 a^2 -2 b^2)\, 
g^2}{4 a\,  b\,  g^2}\,,\label{5fo2}\\
\dot f &=& -\, \fft{( a^2- b^2)\, [ 4  a\,  b\,  f^2\,
 g^2 -  c\, (4 a\,  b\,  c +  a^2\,  f - 
b^2\,  f)\, (1- g^2)]}{16  a^3\,  b^3\,  g^2}\,,\nn\\
\dot g &=& -\, \fft{ c\, (1- g^2)}{4 a\,  b\,  g}\nn
\eea
for the five remaining metric functions.

  Comparing with our results in section (\ref{newg2}), we find that
this conifold-unifying $G_2$ system arises by making the
specialisation $x_1=x_2$ and $y_1=y_2$.  The relations between the two
sets of variables are given by
\bea
&&x_1=x_2= -(a^2 + b^2)\, c\,g\,,\qquad
x_3=(a^2-b^2)\, c + 2a\, b\, f\, g_3\,,\nn\\
&&y_1=y_2=-2a\, b\, c\, f\, g\, (1+g_3)\,,\qquad
y_3=4a^2\, b^2\, g^2\,.\label{xyfromab}
\eea
The first-order equations (\ref{5fo2}) have two simple integration
constants $m$ and $n$, given by
\be
m = (b^2-a^2)\,c\, g^2 + 2a\, b\, f\, g^2\, g_3\,,\qquad
n = (b^2-a^2)\, c + 2 a\, b\, f\,.\label{mncons}
\ee
(The integration constants $m$ and $n$ are called $p$ and $q$ in
\cite{brand,munify}.)  Finally, the constraint (\ref{constr2}) implies
that the Hamiltonian constraint (\ref{hamilton}) is satisfied.

\subsection{Reduction to metrics with $S^3\times T^3$ principal orbits}
\label{abeliansec}

    Recently, a class of $G_2$ metrics with $S^3\times T^3$ principal
orbits was obtained \cite{guyaza}.  This can be seen to arise as a
specialisation of our results in section \ref{newg2} in which a group
contraction of one of the $S^3\sim SU(2)$ factors in the principal
orbits is performed.  It can be seen from (\ref{rhosig}) that a
regular limit of our 3-form $\rho$ and 4-form $\sigma$ will be
obtained if we perform the rescalings
\be
\sigma_i\longrightarrow \lambda\, \sigma_i\,,\quad
x_i\longrightarrow \lambda^{-1}\, x_i\,,\quad
y_i\longrightarrow\, \lambda^{-2}\, y_i\,,\quad
m\longrightarrow \lambda^{-3}\, m\,,
\ee
and then send $\lambda$ to zero.  The metric (\ref{newmet}) reduces to
\bea
ds^2 &=& dt^2 + \fft1{y_1}\, \Big[ x_2\, x_3\, \Sigma_1^2 +
 m\, n\, \Sigma_1\, \sigma_1 +
   m\, x_1 \, \sigma_1^2\Big]\nn\\
&&+\fft1{y_2}\, \Big[ x_3\, x_1\, \Sigma_2^2 +
 m\, n\, \Sigma_2\, \sigma_2 +
   m\, x_2\, \sigma_2^2\Big]\nn\\
&&+\fft1{y_3}\, \Big[x_1\, x_2\, \Sigma_3^2 +
 m\, n \, \Sigma_3\, \sigma_3 +
   m\, x_3 \, \sigma_3^2\Big]\,,\label{s3t3met}
\eea
and the first-order equations (\ref{newfo}) reduce to
\be
\dot x_1 = \sqrt{\fft{y_2\, y_3}{y_1}}\,,\qquad
\dot y_1 = \fft{m\, x_2\, x_3 }{\sqrt{y_1\, y_2\, y_3}}\,,\label{s3t3fo}
\ee
and cyclically for the 2 and 3 directions.  The Hamiltonian constraint
$H=0$ reduces to $m^2\, n^2 - 4m\, x_1\, x_2\, x_3=-4 y_1\, y_2\,
y_3$.  This reproduces the metric and first-order equations obtained
in \cite{guyaza}.  The 3-form and 4-form defined in (\ref{rhosig}) now
become
\bea
\rho &=& n\, \Sigma_1\, \Sigma_2\, \Sigma_3 - m\, \sigma_1\, \sigma_2\,
\sigma_3 + [x_1\, \sigma_1\, \Sigma_2\, \Sigma_3 + 
\hbox{2 cyclic terms}]\,,\nn\\
\sigma &=& y_1\, \sigma_2\, \Sigma_2 \, \sigma_3\, \Sigma_3 +
y_2\, \sigma_3\, \Sigma_3 \, \sigma_1\, \Sigma_1 +
y_3\, \sigma_1\, \Sigma_1 \, \sigma_2\, \Sigma_2\,,\label{rhosigabe}
\eea

   It was shown in \cite{guyaza} that the first-order equations for
these metrics with $S^3\times T^3$ principal orbits can be solved
completely.  Although, as we shall discuss later, no complete and
regular metrics can be obtained (apart from the direct sum of
Eguchi-Hanson and a 3-torus, or flat $\R^4$ and a 3-torus), it is
nevertheless of considerable interest that one can solve the
first-order equations fully \cite{guyaza} in this case.  It provides a
motivation for considering more general possibilities for
group-contraction limits of the metrics obtained in section
\ref{newg2}, and it is to this topic that we move next.

\section{Group contraction limits}
\label{groupcontr}

\subsection{$SU(2)$ group contractions}

     There are three contractions of the $SU(2)$ algebra $d\sigma_i =
-\ft12 \ep_{ijk}\, \sigma_j\wedge \sigma_k$ of left-invariant
differential forms.  In increasing order of degeneracy, they are
\bea
\hbox{\underline{Euclidean group}:}&& \sigma_1\longrightarrow \lambda\,
\sigma_1\,,\qquad \sigma_2\longrightarrow \lambda\, \sigma_2
\,,\qquad \sigma_3\longrightarrow \sigma_3\,,\nn\\
&& d\sigma_1 = -\sigma_2\wedge \sigma_3\,,\qquad
d\sigma_2=\sigma_1\wedge \sigma_3\,,\qquad d\sigma_3=0\,,\label{euclid}\\
&&\nn\\
\hbox{\underline{Heisenberg group}:}&& \sigma_1\longrightarrow \lambda\,
\sigma_1\,,\qquad 
\sigma_2\longrightarrow\, \lambda\, \sigma_2\,,\qquad
\sigma_3\longrightarrow \lambda^2\, \sigma_3\,,\nn\\
&& d\sigma_1 =0\,,\qquad
d\sigma_2= 0\,,\qquad d\sigma_3=-\sigma_1\wedge \sigma_2\,,
\label{heisenberg}\\
&&\nn\\
\hbox{\underline{Abelian group}:}&& \sigma_1\longrightarrow \lambda\,
\sigma_1\,,\qquad \sigma_2\longrightarrow \lambda\, \sigma_2\,,\qquad
\qquad \sigma_3\longrightarrow \lambda\, \sigma_3\,,\nn\\
&& d\sigma_1 =0\,,\qquad
d\sigma_2= 0\,,\qquad d\sigma_3=0\,,\label{abel}
\eea
where $\lambda$ is sent to zero in each case.  Note that the
Heisenberg contraction can be viewed as a further contraction of the
Euclidean group (with an appropriate relabelling of the indices), and
the Abelian group is a further contraction of this.  

   In what follows, we shall consider various group contractions of
$G_2$ metrics with $S^3\times S^3$ principal orbits.  First, we shall
consider the Heisenberg and Euclidean-group contractions of the
general class of $G_2$ metrics that we obtained in section
(\ref{newg2}).  Then, in subsequent subsections, we shall consider in
more explicit detail the group contractions of the metrics obtained in
\cite{munify}, in which there is an additional $U(1)$ factor in the
isometry group.  One can apply the group contractions to one or both
of the $S^3$ factors in the $S^3\times S^3$ principal orbits.  We
shall begin by considering the case where just one of the 3-spheres is
contracted

\subsection{Heisenberg and Euclidean-group contractions}

   In this section, we perform Heisenberg and Euclidean-group
contractions of the new metrics found in section \ref{newg2},
analogous to the Abelian group contraction that we described in
section \ref{abeliansec}.

\bigskip\bigskip
\noindent{\underline{\bf Heisenberg contraction}:}
\bigskip

     The Heisenberg contraction is given by (\ref{heisenberg}).  In
order for the 3-form and 4-form in (\ref{rhosig}) and the metric in
(\ref{newmet}) to have non-singular limits, the following scalings
should also be performed
\bea
&&x_1\longrightarrow \lambda^{-1}\, x_{1}\,,\quad x_2\longrightarrow
\lambda^{-1}\, x_2\,,\quad
x_3\longrightarrow \lambda^{-2}\,x_3\,,\quad
m\longrightarrow \lambda^{-4}\, m\,,\nn\\
&&
y_{1}\longrightarrow \lambda^{-3}\, y_{1}\,,\qquad
y_{2}\longrightarrow \lambda^{-3}\, y_{2}\,,\qquad
y_3\longrightarrow \lambda^{-2}\,y_3\,,
\eea
while $n$ is unscaled. The forms defined in (\ref{rhosig}) now become
\bea
\rho &=& n\, \Sigma_1\, \Sigma_2\, \Sigma_3 - m\, \sigma_1\, \sigma_2\,
\sigma_3 + x_1\, \sigma_1\, \Sigma_2\, \Sigma_3 + 
+ x_2\, \sigma_2\, \Sigma_3\, \Sigma_1 +
x_3\, (\sigma_3\, \Sigma_1\, \Sigma_2 - \sigma_1\,\sigma_2\,
\Sigma_3)\,,\nn\\
\sigma &=& y_1\, \sigma_2\, \Sigma_2 \, \sigma_3\, \Sigma_3 +
y_2\, \sigma_3\, \Sigma_3 \, \sigma_1\, \Sigma_1 +
y_3\, \sigma_1\, \Sigma_1 \, \sigma_2\, \Sigma_2\,,\label{rhosighei}
\eea
and the metric (\ref{newmet}) becomes
\bea
ds^2 &=& dt^2 + \fft1{y_1}\, \Big[ x_2\, x_3\, \Sigma_1^2 +
 (m\, n - x_3^2)\, \Sigma_1\, \sigma_1 +
   m\, x_1\, \sigma_1^2\Big]\nn\\
&&+\fft1{y_2}\, \Big[ x_3\, x_1\, \Sigma_2^2 +
 (m\, n - x_3^2)\, \Sigma_2\, \sigma_2 +
   m\, x_2\, \sigma_2^2\Big]\nn\\
&&+\fft1{y_3}\, \Big[ (n\, x_3 + x_1\, x_2)\, \Sigma_3^2 +
 (m\, n + x_3^2)\, \Sigma_3\, \sigma_3 +
   m\, x_3\, \sigma_3^2\Big]\,,\label{newmethei}
\eea
The first-order equations after taking this Heisenberg scaling limit
will be
\bea
&&\dot x_1 = \sqrt{\fft{y_2\,y_3}{y_1}}\,,\qquad
\hbox{cyclically for 2 and 3 directions}\,,\nn\\
&& \dot y_1 = \fft{m\, x_2\, x_3}{\sqrt{y_1\, y_2\, y_3}}\,,\qquad
\dot y_2 =\fft{m\, x_1\, x_3}{\sqrt{y_1\, y_2\, y_3}}\,,\qquad
\dot y_3 = \fft{m\, n\, x_3 + m\, x_1\, x_2- x_3^3}{
\sqrt{y_1\, y_2\, y_3}}\,,
\eea
with the Hamiltonian constraint $H=0$ giving
\be
m^2\, n^2\, -2m\,n\, x_3^2 - 4m\, x_1\, x_2\, x_3 +
x_3^4 + 4 y_1\, y_2\, y_3=0\,.
\ee

\bigskip\bigskip
\noindent{\underline{\bf Euclidean contraction}:}
\bigskip

      The Euclidean contraction is given by (\ref{euclid}).  To
obtain a non-singular system in this limt, we need the following
scalings:
\bea
&&x_{1}\longrightarrow \lambda^{-1}\, x_{1}\,,\quad
x_{2}\longrightarrow \lambda^{-1}\, x_{2}\,,\quad
m\longrightarrow \lambda^{-2}\, m\,,\nn\\
&&
y_{1}\longrightarrow \lambda^{-1}\, y_{1}\,,\quad
y_{2}\longrightarrow \lambda^{-1}\, y_{2}\,,\quad
y_3\longrightarrow \lambda^{-2}\, y_3\,,
\eea
while $x_3$ and $n$ are unscaled.
The forms $\rho$ and $\sigma$ defined in (\ref{rhosig}) now become
\bea
\rho &=& n\, \Sigma_1\, \Sigma_2\, \Sigma_3 - m\,\sigma_1\,\sigma_2\,
\sigma_3 + x_1\, (\sigma_1\, \Sigma_2\, \Sigma_3 -
\sigma_2\, \sigma_3\, \Sigma_1)\nn\\
&&+ x_2\, (\sigma_2\, \Sigma_3\, \Sigma_1 -
\sigma_3\, \sigma_1\, \Sigma_2) + x_3\, \sigma_3\, \Sigma_1\,
\Sigma_2\,,\nn\\
\sigma &=& y_1\, \sigma_2\, \Sigma_2 \, \sigma_3\, \Sigma_3 +
y_2\, \sigma_3\, \Sigma_3 \, \sigma_1\, \Sigma_1 +
y_3\, \sigma_1\, \Sigma_1 \, \sigma_2\, \Sigma_2\,,\label{rhosigeuc}
\eea
and the metric (\ref{newmet}) becomes
\bea
ds^2 &=& dt^2 + \fft1{y_1}\, \Big[ (n\, x_1 + x_2\, x_3)\, \Sigma_1^2 +
 (m\, n + x_1^2 - x_2^2)\, \Sigma_1\, \sigma_1 +
   m\, x_1\, \sigma_1^2\Big]\nn\\
&&+\fft1{y_2}\, \Big[ (n\, x_2 + x_3\, x_1)\, \Sigma_2^2 +
 (m\, n + x_2^2 - x_1^2)\, \Sigma_2\, \sigma_2 +
   m\, x_2\, \sigma_2^2\Big]\nn\\
&&+\fft1{y_3}\, \Big[ x_1\, x_2\, \Sigma_3^2 +
 (m\, n - x_1^2 - x_2^2)\, \Sigma_3\, \sigma_3 +
   (m\, x_3 + x_1\, x_2)\, \sigma_3^2\Big]\,,\label{yaumet}
\eea
The first-order equations in this limit will be
\bea
&&\dot x_1 = \sqrt{\fft{y_2\,y_3}{y_1}}\,,\qquad
\hbox{and cyclically for 2 and 3 directions}\,,\nn\\
&&\dot y_1 =\fft{m\, n\, x_1 + m\, x_2\, x_3 + x_1\, (x_2^2 - x_1^2)}{
\sqrt{y_1\, y_2\, y_3}}\,,\quad
\dot y_2 = \fft{m\, n\, x_2 + m\, x_1\, x_3 + x_2\,
(x_1^2-x_2^2)}{\sqrt{y_1\, y_2\, y_3}}\,,\nn\\
&&\dot y_3 = \fft{m\, x_1\, x_2}{\sqrt{y_1\, y_2\, y_3}}\,,
\eea
with the Hamiltonian constraint $H=0$ giving
\be
m^2\, n^2 - 2m\, n\, (x_1^2 + x_2^2) - 4m\, x_1\, x_2\, x_3\,
+ (x_1^2 - x_2^2)^2 + 4 y_1\, y_2\, y_3 =0\,.
\ee

   It should be noted that in all three of the three scalings of the
metrics in section \ref{newg2}, \ie the Abelian, Heisenberg and
Euclidean cases, the coefficient $m$ is rescaled while the coefficient
$n$ is not.  Thus one cannot take such contraction limits in the case
where $m=n$, which, as we showed in section \ref{sixfnsec}, reduces to
the six-function system found in \cite{cglp6fun,bragugogo}.  On the
other hand, we {\it can} take these contraction limits in the other
specialisation that we discussed in (\ref{conifoldsec}), where there
is an additional $U(1)$ in the isometry group and the metrics are
described by the first-order system obtained in \cite{munify}.

   We shall now turn to a more detailed investigation of the various
group contractions for the $G_2$ metrics found in \cite{munify}.

\section{The contraction from $S^3\times S^3$ with 
$U(1)$ isometry}\label{abeliansec1}

\subsection{$S^3\times T^3$ principal orbits}\label{s3t3u1}

   In the case of the $G_2$ metrics with an additional $U(1)$ isometry
in the principal $S^3\times S^3$ orbits, it is possible to study the
contraction limits in a more explicit manner.  The extra $U(1)$
isometry arises if we set the functions $x_i$ and $y_i$ in two of the
three $S^3$ directions equal, for example $x_1=x_2$, $y_1=y_2$.  As we
showed in section \ref{conifoldsec}, the new $G_2$ metrics in section
\ref{newg2} are then equivalent to those found in \cite{munify}, which
are given in equations (\ref{d7ans2}), together with the conditions
(\ref{constr2}) and (\ref{5fo2}) for $G_2$ holonomy.  For our present
purposes, we find it more convenient to work with the metrics written
as in (\ref{d7ans2}), which we therefore take as our starting point
for studying the contraction limits. (The solutions that we obtain in
this section are contained, albeit in a different parameterisation,
within the general solutions constructed in \cite{guyaza}.)

   We now make the rescaling as in (\ref{abel}), at the same time 
rescaling the metric functions in (\ref{d7ans2}) according to
\be
g\longrightarrow \fft{g}{\lambda}\,,\qquad g_3\longrightarrow
\fft{g_3}{\lambda}\,.\label{rescale1}
\ee
The metric (\ref{d7ans2} then becomes
\bea
ds_7^2 &=& dt^2 +  a^2\, [(\Sigma_1 +  g\, \a_1)^2 +
                          (\Sigma_2 +  g\, \a_2)^2]
+  b^2\,  [(\Sigma_1 -  g\, \a_1)^2 +
                          (\Sigma_2 -  g\, \a_2)^2]\nn\\
&&+  c^2\, \Sigma_3^2 +  f^2\, (\Sigma_3 +  g_3\,
\a_3)^2\,.\label{d7ans3}
\eea
The algebraic constraint (\ref{constr2}) becomes
\be
2 a\, b\, f = c\, (b^2-a^2)\,,\label{constr3}
\ee
while the five first-order equations (\ref{5fo2}) become
\bea
\dot a &=& \fft{ 4a^4 - 4 a^2\, b^2 - 3 a^2\, c^2 
        - b^2\, c^2}{16 a^2\,  b\,  c}\,,\nn\\
\dot b &=& \fft{ 4 b^4 - 4 a^2\, b^2 - a^2\, c^2 
             - 3 b^2\, c^2}{16  a\,  b^2\,  c}\,,\nn\\
\dot c &=& \fft{ c^2 -2a^2 -2 b^2}{4 a\,  b}\,,\label{5fo3}\\
\dot g &=&  \fft{ c\, g}{4 a\,  b}\,, \qquad \dot g_3 = -\fft{c\,
g_3}{2 a\, b}\,.
\nn
\eea
Note that the last two equations imply $g^2\, g_3=$constant.

   In this contraction limit the relations (\ref{xyfromab}) that
we obtained by matching the specialisation of section \ref{newg2} 
where $x_1=x_2$ and $y_1=y_2$ to the metrics in \cite{munify} 
can be inverted simply, to give
\bea
&&a^2= \fft{2 x_1\, x_3 + n\, \sqrt{-m\, x_3}}{4 y_1}\,,\qquad
b^2 =  \fft{2 x_1\, x_3 - n\, \sqrt{-m\, x_3}}{4 y_1}\,,\qquad
g^2 = -\fft{m}{x_3}\,,\nn\\
&&c^2 = \fft{4 x_1^2\, x_3 + m\, n^2}{4 x_3\, y_3}\,,\qquad
f^2 = -\fft{m\, n^2}{4 x_3\, y_3}\,,\qquad
g_3 = -\fft{2 x_3}{n}\,.
\eea
Note that the constraint (\ref{constr3}) is indeed satisfied, and
also that
\be
g^2\, g_3 =\fft{2m}{n}\,.
\ee
   From (\ref{ysol}), we also see that $c\, (b^2-a^2)$ is a constant,
and so from (\ref{constr3}) we have $a\, b\, f=$constant.  In terms of
the metric variables in \cite{guyaza}, this translates into $m^2\, n^2
+ 4m\, x_1^2\, x^3 = - 4y_1^2\, y_3$, which is indeed the condition of
the vanishing of the Hamiltonian $H$ given in their eq (4.32).

    It should be remarked that whilst it is straightforward to take
the limit $\lambda\longrightarrow 0$ in order to get the first-order
equations for $a$, $b$, $c$ and $g$, one has to be slightly more
careful in order to get the algebraic constraint (\ref{constr3}) and
the first-order equation for $g_3$.  Specifically, (\ref{constr3}) is
obtained by solving (\ref{constr2}) for $f$ and then taking the limit
$\lambda\longrightarrow 0$.  The first-order equation for $g_3$ is
obtained by differentiating (\ref{constr2}), and then using
(\ref{constr2}) itself to substitute for $f$ in the resulting
expression for $\dot g_3$.  This gives
\be
\dot g_3 = -\fft{c\, (g^2-g_3)(1+g_3)}{2a\, b\, g^2}\,.\label{g3dot}
\ee
Finally, after making the replacements $g\longrightarrow g/\lambda$
and $g_3\longrightarrow g_3/\lambda$, we obtain the equation for $\dot
g_3$ in (\ref{5fo3}).  It is worth noting also that another way of
obtaining (\ref{constr3}) and (\ref{5fo3}) is by directly imposing
$d\Phi_3=0$ and $d{*\Phi_3}=0$, where the vielbein in
(\ref{special3form}) is the natural one read off from (\ref{d7ans3}).

\subsubsection{Solving the equations for $S^3\times T^3$ with 
$U(1)$ isometry}

  Introduce a new radial variable $r$ by $dr=\ft12 a\, b\, c\, dt$,
and define $A\equiv a^2$, $B\equiv b^2$, $C\equiv c^2$.  The equations
for $\dot a$, $\dot b$ and $\dot c$ in (\ref{5fo3}) become
\bea
A' &=& \fft{A}{B\, C} - \fft{1}{C} - \fft{3}{4B} - \fft1{4A}\,,\nn\\
B' &=& \fft{B}{A\, C} - \fft{1}{C} - \fft{3}{4A} - \fft1{4B}\,,\nn\\
C' &=& -\fft{2}{A} -\fft{2}{B} + \fft{C}{A\, B}\,.\label{abceq}
\eea
Now define
\be
A=X+Y\,,\qquad B=X-Y\,,
\ee
and introduce a new radial variable $\rho$ such that $dr=-A\, B\, d\rho$.
The equations (\ref{abceq}) become
\be
\fft{dX}{d\rho} = X- \fft{2Y^2}{C} \,,\qquad \fft{dY}{d\rho} 
= \ft12 Y -\fft{2 X\, Y}{C}\,,\qquad
\fft{dC}{d\rho}  = 4 X -C \,.\label{xyceq}
\ee
Note that we can deduce from these that
\be
C\, Y^2 =k^2\,,\label{ysol}
\ee
where $k$ is a constant.  (We must have $k^2$ non-negative, since
$C=c^2$ is non-negative.) 

    Equation (\ref{ysol}) implies that if $k$ is non-vanishing, the
metrics will be singular.  This follows from (\ref{constr3}), 
since we therefore have
\be
2 a\, b\, f = c\, (b^2-a^2) =k\,.
\ee
This shows that at any putative short-distance endpoint, which would
be characterised by a smooth degeneration of a circle or sphere to the
origin of (spherical) polar coordinates, some other metric function
would be diverging there.  This is a rather common feature of
Ricci-flat metrics with principal orbits that contain torus factors;
various examples with special holonomy were discussed in
\cite{gilupost}.  Indeed, we shall show in section \ref{s3t3sing} that
aside from the direct product of Eguchi-Hanson and $T^3$, there are no
other non-trivial regular solutions to the $G_2$ metrics with
$S^3\times T^3$ principal orbits that were obtained in \cite{guyaza}.
 
   It is, nevertheless, of interest to study the explicit solutions
for the $G_2$ metrics in section \ref{s3t3u1}. In particular, we shall
see how the Eguchi-Hanson times $T^3$ metric emerges as a non-singular 
Gromov-Hausdorff limit of a family of singular $G_2$ metrics, in which
$k$ goes to zero.

   Solving the $dC/d\rho$ in equation (\ref{xyceq} for $X$, and
plugging into the $dX/d\rho$ equation, also making use of
(\ref{ysol}), we therefore obtain the following second-order equation
for $C$:
\be
\fft{d^2 C}{d\rho^2} - C + \fft{8k}{C^2}=0\,.\label{ceq}
\ee
Multiplying by $dC/d\rho$, this can be integrated once, giving
\be
\fft{dC}{d\rho} = \sqrt{C^2 + 16 k^2\, C^{-1} + \mu}\,,\label{cfo}
\ee
where $\mu$ is a constant.
Note that we have chosen the positive square root here, 
because the $dC/d\rho$ equation in (\ref{xyceq}) gives
$X=\ft14(C+dC/d\rho)$, and hence
\be
X= \ft14 C + \ft14 \sqrt{C^2 + 16 k^2\, C^{-1} + \mu}\,.\label{xeq}
\ee
This would have been negative at small $C$ if we had chosen the other
sign in (\ref{cfo}), contradicting the fact that
$X=\ft12(A+B)=\ft12(a^2+b^2)$ is non-negative.  Our choice here is
adapted to allowing $c$ to become small.

    Substituting from (\ref{xeq}) and (\ref{ysol}), we therefore find
that
\be
A= \ft14 C + \ft14\sqrt{C^2 + 16k^2\, C^{-1} + \mu} + \fft{k}{\sqrt
C}\,,\qquad
B= \ft14 C + \ft14 \sqrt{C^2 + 16k^2\, C^{-1} + \mu} -\fft{k}{\sqrt
C}\,.\label{absol}
\ee
Integrating (\ref{cfo}), we obtain
\be
\rho = \int_0^C\fft{dx}{\sqrt{x^2 + 16k^2\, x^{-1} + \mu}}\,.
\label{cint}
\ee
We have chosen the integration limit so that $C$ vanishes at $\rho=0$.

     The integral can be evaluated explicitly, if rather opaquely, in
terms of elliptic functions.  It is convenient to use $C$ as the
radial variable.  We have $dt=2/(a\, b\, c)\, dr= -2(a\, b/c)\,
d\rho$, and so
\be
dt^2 = \fft{4 A\, B}{C}\, d\rho^2= \fft{4 A\, B\, dC^2}{
 C^2 + 16k^2\, C^{-1} + \mu} \,.
\ee
Thus if we let $C\equiv z^2$, we get the metric in the form
\bea
ds^2 &=& \fft{16 A\, B\, dz^2}{z^4 + 16 k^2/z^2 + \mu} + A\, 
[(\Sigma_1+g\, \a_1)^2 + (\Sigma_2+g\, \a_2)^2] \nn\\
&& +  B\, 
[(\Sigma_1 - g\, \a_1)^2 + (\Sigma_2- g\, \a_2)^2]
+ z^2\, \Sigma_3^2 + \fft{4k^2}{A\, B}\, (\Sigma_3-g_3\, \a_3)^2\,,
\label{abelianmet}
\eea
with
\be
A= \ft14 z^2 + \ft14\sqrt{z^4 + 16k^2\, z^{-2} + \mu} + 
\fft{k}{z}\,,\qquad
B= \ft14 z^2 + \ft14 \sqrt{z^4 + 16k^2\, z^{-2} + \mu} -\fft{k}{z}
\,.\label{absol2}
\ee
The functions $g$ and $g_3$ are then given by
\be
g= g_0\, e^{-\fft12\rho}\,,\qquad g_3= \tilde g_0\, e^{\rho} 
\label{gsols}
\ee
where $\rho$ is given by (\ref{cint}) with $C=z^2$, and $g_0$ and $\td
g_0$ are constants.  At $z\rightarrow \infty$, the metric becomes
locally $\R^4\times T^3$.  For a non-vanishing value of $k$, the
metric has a singularity at $z=0$, where either $A$ or $B$ becomes
divergent.  If we had chosen the negative root of (\ref{ceq}) for
$dC/d\rho$, the singularity would occur at some finite $z_0$ where $A$
or $B$ vanishes.

    Note that if we take $k$ to vanish, then the integral in
(\ref{cint}) becomes elementary, allowing us to obtain simple formulae
for $g$ and $g_3$.  Letting $\mu=\ell^4$ and introducing a new radial
variable $y$ defined by $y^2=z^2+\sqrt{z^4+\ell^4}$, we find $\rho=2
\log(y/\ell)$.  The metric (\ref{abelianmet}) then becomes
\be
ds^2 = 2\Big(1-\fft{\ell^4}{y^4}\Big)^{-1}\, dy^2 + \ft12 y^2\, 
\Big(1-\fft{\ell^4}{y^4}\Big)\, \Sigma_3^2 + \ft12 y^2\, (\Sigma_1^2
+\Sigma_2^2) + g_0^2\, \ell^2\, (\a_1^2+\a_2^2) + \hat g_0^2\, \ell^2\, 
\a_3^2\,,\label{eht3}
\ee
where we have replaced the constant $\td g_0$ in (\ref{gsols}) by
$\hat g_0= 8 k\, \td g_0/\ell^3$ before sending $k$ to zero.  The
metric (\ref{eht3}), which is the Gromov-Hausdorff limit of the
general solution (\ref{abelianmet}) (with the radius of the circle
described by $\a_3$ sent to zero), is nothing but the direct sum of
the Eguchi-Hanson metric and $T^3$.  As was shown in \cite{begipapo},
the Eguchi-Hanson metric is asymptotic to $\R^4/\bZ_2$, and so the
regular 7-metric we obtain here in the Gromov-Hausdorff limit $k=0$ is
asymptotic to $\R^4/\bZ_2\times T^3$.  It should be emphasised,
however, that before the Gromov-Hausdorff limit is taken, the metrics
are singular.

   The integration (\ref{cint}) again becomes elementary if $\mu$ is
set to zero instead of $k$, yielding
\be
\rho= -\ft23 \log(4k) + \ft23 \log(z^3+\sqrt{z^6+16k^2})\,.
\ee
If we now introduce a new radial coordinate $y$ such that $y^3=
z^3 +  \sqrt{z^6+16k^2}$, the metric (\ref{abelianmet}) becomes
\bea
ds^2&=& 2^{4/3}\Big(1-\fft{16 k^2}{y^6}\Big)^{-1/3}\,  dy^2 
+ 2^{-2/3} y^2\, \Big(1-\fft{16 k^2}{y^6}\Big)^{2/3}\, \Sigma_3^2
 \nn\\
&&+ 32 2^{1/3} k^2\, y^{-4}\, \Big(1-\fft{16 k^2}{y^6}\Big)^{-1/3}\, 
(\Sigma_3 + \td g_0\, y^2\, \a_3)^2\nn\\
&& 
+ 2^{-5/3} y^2\,\Big(1-\fft{16 k^2}{y^6}\Big)^{-1/3}\, 
\Big(1+ \fft{4k}{y^3}\Big)\, [(\Sigma_1+ g_0\,y^{-1}\, \a_1)^2 +
   (\Sigma_2+ g_0\,y^{-1}\, \a_2)^2]\nn\\
&&+ 2^{-5/3} y^2\,\Big(1-\fft{16 k^2}{y^6}\Big)^{-1/3}\, 
\Big(1- \fft{4k}{y^3}\Big)\, [(\Sigma_1- g_0\,y^{-1}\, \a_1)^2 +
   (\Sigma_2- g_0\,y^{-1}\, \a_2)^2]\,,
\eea
after rescaling constants $g_0$ and $\td g_0$.  This is, as expected,
as singular metric.  The metric runs from (locally) $\R^4\times T^3$
at $y=\infty$ to a singularity at $y=(4k)^{1/3}$.

\subsection{$S^3\times\hbox{(Heisenberg)}$ principal orbits}

    Here, we instead perform the Heisenberg contraction given in
(\ref{heisenberg}).  One must now perform the associated metric rescalings 
\be
g\longrightarrow \fft{g}{\lambda}\,,\qquad g_3 \longrightarrow 
 \fft{g_3}{\lambda^2}\,,
\ee
giving the metric
\bea
ds_7^2 &=& dt^2 +  a^2\, [(\Sigma_1 +  g\, \beta_1)^2 +
                          (\Sigma_2 +  g\, \beta_2)^2]
+  b^2\,  [(\Sigma_1 -  g\, \beta_1)^2 +
                          (\Sigma_2 -  g\, \beta_2)^2]\nn\\
&&+  c^2\, \Sigma_3^2 +  f^2\, (\Sigma_3 +  g_3\,
\beta_3)^2\,.\label{d7ans4}
\eea
It is evident from (\ref{constr2}) and (\ref{5fo2}) that the
constraint will now become
\be
 g_3 =  \fft{ [2a\, b\, f-c\, (b^2-a^2)]\, g^2}{2 a\,  b\,  f}\,,
\label{constr4}
\ee
while the first-order equations for $(a,b,c,f,g)$ will become
\bea
\dot a &=& \fft{4 a^2\, ( a^2- b^2)-  c^2\, (5  a^2- b^2) - 4
a\,  b\,  c\,  f}{16 a^2\,  b\,  c}\,,\nn\\
\dot b &=& -\, \fft{4  b^2\,
( a^2 - b^2) + c^2\, (5 b^2 -  a^2) - 4 a\,  b\,
 c\,  f}{16  a\,  b^2\,  c}\,,\nn\\
\dot c &=& \fft{ c^2 -2a^2 -2 b^2}{4 a\,  b}\,,\label{5fo4}\\
\dot f &=& -\fft{(a^2-b^2)\, [4a\, b\, (c^2+f^2) + c\, f\,
(a^2-b^2)]}{16 a^3\, b^3}\,,\nn\\
\dot g &=&  \fft{ c\, g}{4 a\,  b}\,.
\nn
\eea
This first-order system has two integration constants $m=2a\, b\, f\,
g^2\, g_3$ and $n=(b^2-a^2)\, c + 2 a\, b\, f$.  We can use the
constant $n$ to express the function $f$ such that the functions $a$,
$b$ and $c$ form a closed first-order system.  We have not obtained
the general solution of these equations.

\subsection{$S^3\times\hbox{(Euclidean)}$ principal orbits}

   Here, we instead perform the contraction (\ref{euclid}) of the
$SU(2)$ algebra for the left-invariant 1-forms $\sigma_i$ in 
(\ref{d7ans2}).  Correspondingly, we now rescale only the metric
function $g$, according to $g\longrightarrow g/\lambda$, giving
\bea
ds_7^2 &=& dt^2 +  a^2\, [(\Sigma_1 +  g\, \gamma_1)^2 +
                          (\Sigma_2 +  g\, \gamma_2)^2]
+  b^2\,  [(\Sigma_1 -  g\, \gamma_1)^2 +
                          (\Sigma_2 -  g\, \gamma_2)^2]\nn\\
&&+  c^2\, (\Sigma_3-\gamma_3)^2 +  f^2\, (\Sigma_3 +  g_3\,
\gamma_3)^2\,.\label{d7ans5}
\eea
The first-order equations for $G_2$ holonomy now become

\bea
\dot a &=& \fft{ 4a^4 - 4 a^2\, b^2 - 3 a^2\, c^2 
        - b^2\, c^2}{16 a^2\,  b\,  c}\,,\nn\\
\dot b &=& \fft{ 4 b^4 - 4 a^2\, b^2 - a^2\, c^2 
             - 3 b^2\, c^2}{16  a\,  b^2\,  c}\,,\nn\\
\dot c &=& \fft{ c^2 -2a^2 -2 b^2}{4 a\,  b}\,,\label{5fo5}\\
\dot g &=&  \fft{ c\, g}{4 a\,  b}\,, \qquad \dot g_3 = -\fft{c\,
(g_3+1)}{2 a\, b}\,,\nn
\nn
\eea
together with the same algebraic constraint (\ref{constr3}) as in the
Abelian contraction:
\be
2a\, b\, f = c\, (b^2-a^2)\,.\label{constr6}
\ee
Note that the last two equations in (\ref{5fo5}) imply $g^2\,
(g_3+1)=$constant.  Again, this first-order system has two integration
constants $m$ and $n$ given by (\ref{mncons}). The first three
equations in (\ref{5fo5}) are the same as in the Abelian contraction.
(The slightly delicate procedure for taking the limit to get the
constraint (\ref{constr6}) and the first-order equation for $\dot g_3$
goes in the same way as we described in section \ref{abeliansec} for
the Abelian limit.  It can be seen that (\ref{g3dot}) now gives the
expression for $\dot g_3$ appearing in (\ref{5fo5}).)

\section{Global considerations in the $S^3\times T^3$ metrics of 
$G_2$ holonomy}\label{s3t3sing}

   In this section, we make some observations about solutions of the
system of $G_2$ metrics with $S^3\times T^3$ principal orbits that was
obtained in \cite{guyaza}, and which is reproduced in section
\ref{abeliansec}.  In particular, we shall present simple arguments
which show that there can be no complete and regular metrics within
this class, other than flat $\R^4\times T^3$, or the direct product of
Eguchi-Hanson and $T^3$.

    It was observed in \cite{guyaza} that measured in the $G_2$ metric
(\ref{s3t3met}), the volumes of the $S^3$ and $T^3$ factors in the
principal orbits are bounded below, with
\be
\hbox{Vol}(S^3) \ge |n|\,,\qquad \hbox{Vol}(T^3)\ge |m|\,.
\label{bounds}
\ee
(These results can easily be seen by using the Hamiltonian constraint
in the expressions $(x_1 x_2 x_3)\, (y_1 y_2 y_3)^{-1/2}\,
\Sigma_1\Sigma_2\Sigma_3$ and $m^{3/2}\,
(x_1x_2x_3/(y_1y_2y_3))^{1/2}\, \sigma_1\sigma_2\sigma_3$ for the
volume forms that one can read off from (\ref{s3t3met}).)  It follows,
therefore, that if $m\, n\ne 0$, the principal orbits will never
collapse, for any value of $t$.  Under these circumstances, one can
never obtain a complete and regular metric, since there will be no
short-distance endpoint at which the metric ``closes off.'' Thus the
radial coordinate would be running between two endpoints corresponding
to asymptotic infinities, but by a standard theorem one can have at
most one asymptotic infinity in a complete regular Ricci-flat metric.

   The only possibility for regular metrics, therefore, is to have
$m\, n=0$, allowing one or other of the $S^3$ or $T^3$ to collapse on
singular orbits.  For example, the $S^3$ might degenerate to $S^2$ on
such an orbit, which could then give rise to an $S^2\times T^3$
``bolt'' at short distance, closing off the metric.  However, from
(\ref{s3t3met}) we see that if $m\, n=0$, the metric becomes purely
diagonal,
\be
ds^2 = dt^2 + \fft{x_2\, x_3}{y_1}\, \Sigma_1^2 + 
\fft{x_3\, x_1}{y_2}\, \Sigma_2^2 + \fft{x_1\, x_2}{y_3}\, \Sigma_3^2
+ \fft{m\, x_1}{y_1}\, \sigma_1^2 +  \fft{m\, x_2}{y_2}\, \sigma_2^2+
 \fft{m\, x_3}{y_3}\, \sigma_3^2
\,,\label{summet}
\ee
with the Hamiltonian constraint becoming $y_1 y_2 y_3 = m\, x_1 x_2 x_3$.

   If the $T^3$ remains uncollapsed ($m\ne 0$), we can perform a
Kaluza-Klein reduction on the $T^3$.  The absence of off-diagonal
terms in (\ref{summet}) means that there will be no Kaluza-Klein
vectors, and so the Ricci-flatness of (\ref{summet}) will translate,
in the reduced $D=4$ equations, to a system of equations that includes
three ``dilaton equations,'' each of the form
\be
\square \, \phi = 0\,,\label{dilatoneq}
\ee
where we have a dilaton $\phi\sim \log(x_i/y_i)$ for each reduction
circle.  Since the metric is of cohomogeneity one, this means that
$\phi$ is a function only of the radial coordinate $t$, and so
(\ref{dilatoneq}) is just $d(\sqrt{g}\, \dot\phi)/dt=0$, where $g$ is
the determinant of the reduced 4-metric. The general solution is
\be
\phi = c_1 + c_2\, \int^t \fft{dt'}{\sqrt{g(t')}}\,,\label{phisol}
\ee
where $c_1$ and $c_2$ are constants.  In a putative regular 7-metric
the radius of each circle within $T^3$ remains non-vanishing at
short-distance, and so $\phi$ is finite there.  If a $q$-sphere within
the $S^3$ collapses, for any $1\le q\le 3$, we will have
$\sqrt{g(t)}\sim t^{q}$, and so in order to have $\phi$ finite at
short distance (\ie at $t=0$) it must be that $c_2=0$, and hence we
have $\phi=\,$constant.  Repeating for all three dilatons, we conclude
that $x_i/y_i$ is a constant for each $i$.  Without loss of generality
we can rescale the lengths of the circles so that $x_i=y_i$ (and hence
$m=1$), and so the metric (\ref{summet}) then becomes simply
\be
ds^2 = dt^2 + \fft{y_2\, y_3}{y_1}\, \Sigma_1^2 +  
\fft{y_3\, y_1}{y_2}\, \Sigma_2^2 +  \fft{y_1\, y_2}{y_3}\, \Sigma_3^2 
  + d\theta_1^2 + d\theta_2^2 + d\theta_3^2\,,
\ee
where we write the $T^3$ 1-forms as $\sigma_i=d\theta_i$.  
The first-order equations (\ref{s3t3fo}) then become
\be
\dot y_1 = \sqrt{\fft{y_2\, y_3}{y_1}}\,, \qquad \hbox{and cyclic}\,.
\ee
These are just one of the systems of first-order equations for
Bianchi-IX 4-metrics of $SU(2)$ holonomy, which admit the
Eguchi-Hanson metric as a complete regular solution if two of the
three directions are set equal.  If all directions are unequal, the
solutions are incomplete, with curvature singularities
\cite{begipapo}.  Thus we have established that the only regular
metrics within the class obtained in \cite{guyaza}, with $m\ne0$ so
that the $T^3$ factor does not collapse, are either flat $\R^4$ times
$T^3$, or else the product of Eguchi-Hanson and a flat 3-torus.  In
this latter case, the metric approaches $\R^4/\bZ_2\times T^3$ at
large distance.

\section{A torus splitting theorem} 

    The purpose of this section is to show that the problems found in
section \ref{s3t3sing} when constructing a non-trivial and
non-singular fibration by tori in the specific example of the metrics
in \cite{guyaza} are in fact generic for Ricci-flat metrics, as long
as one supposes the torus action to be by isometries. This has
consequences for implementing at the level of concrete and explicit
exact metrics some of the ideas of \cite{syza} and \cite{guyaza} on
fibrations by special Lagrangian and associative tori
respectively. These difficulties will be illustrated in the following
section by means of concrete examples of exact but incomplete
Calabi-Yau and $G_2$ metrics drawn from earlier work \cite{gilupost}
on contractions. In the $G_2$ case, they are obtained by a more
drastic contraction of the $SU(2)\times SU(2)$ isometry group than to
$SU(2) \times T^3$.

\subsection{Kaluza-Klein reductions}

   We shall begin by reviewing the standard toroidal reduction of a
$(d+k)$-dimensional Riemannian manifold $E$ to a $d$-dimensional
Riemannian base manifold $B$ with metric $g_{\mu \nu}$.  Later, we
shall replace the torus group fibres $T^k$ by a general unimodular Lie
group $G$. If the torus action is free, $B$ will be a smooth manifold.
If the action is not free, $B$ may be singular as a manifold and/or
its metric may be singular. The conclusion of our theorem is that if
the action is free then the metric splits as an unwarped and untwisted
product.  The basic {\sl local} formulae are the metric ansatz:
\be 
ds^2 = \exp{ {2 U \over k}} {\hat h} _{mn} ( dy^m + A^m_\mu\,
d x^\mu ) (dy^n +  A^n_\nu \, dx ^\nu  )\, A \, B + \exp { {2 U \over
d-2}}\,  g_{\mu \nu}\,  dx^\mu\,  dx^\nu, 
\ee
and the $d$-dimensional action from which the Ricci-flat
conditions may be derived.  The Lagrangian density is
\bea {\cal L}&=&\sqrt{-g} \Bigl ( { R } - \ft14 \exp({2kU
\over k-2})\,  {\hat h}_{mn}\,  F^m_{\mu \nu}\, F^n\thinspace^{\mu
\nu} + \ft14 g^{\mu \nu} \, \bigr ( {\rm Tr} \thinspace
 {\hat h} ^{-1} \partial _\mu {\hat h} {\hat h} ^{-1} \partial _\nu
 {\hat h} \bigl )\nn\\
&&+ { } ( { 1\over d-1} + {1 \over k^2 }  )\,
 g^{\mu \nu}\,
\partial_\mu U \partial_\nu U \Bigl ) .
\eea

   The matrix ${\hat h}_{mn}$ has unit determinant, and
\be
F^m_{\mu \nu}= \partial _\mu A^m_\nu- \partial _\nu A^n_\mu.
\ee
Thus the quantity $\exp(U)$ is the volume of the toroidal fibres.

   We  now show that the vector fields must vanish under suitable
global assumptions.  To do so, we note that the equations of motion
of the vectors imply that
\be
\nabla _\mu \, \Big( \exp({2 k U \over k-2})\,  {\hat h}_{mn}\, A^m_\nu
\, F^m \thinspace^{\nu \mu} \Big) = -\ft12 \exp ({
2 kU \over k-2})\,  {\hat h}_{mn}\,  F^m_{\mu \nu} \, F^n \thinspace
^{\mu \nu} \label{current}.
\ee
Assuming that the vector potentials $A^m_\mu$ are globally defined,
then integration over the base, together with the assumption that the
boundary terms at infinity vanish (which is typically the case if the
metric approaches the flat metric on $T^k \times {\Bbb E} ^d$
sufficiently fast), shows that the vectors must vanish.

    Having established the vanishing of the vectors, we can now show
that the volume of the toroidal fibres must be constant.  The quantity
$U$ satisfies
\be 
\nabla^2 U = { k \over 2 \kappa ^2 (k-2) } \exp({2kU \over
k-2})\,
 {\hat h} _{mn}\,  F^m_{\mu \nu} \, F^n\thinspace ^{\mu \nu},
\label{Earnshaw} \ee
We see that even if the vectors are non-vanishing, the volume of
the fibres can have no interior maximum. Moreover by integrating
over the base $B$ of the fibration we see that unless there is a
boundary contribution from infinity, the field strengths must
vanish. In that case, multiplication by $U$ and integration over
the base $B$ then shows that $U$ must be constant.

    Now the action for the scalars ${\hat h}_{mn}$ reduces to a
harmonic map into the space of unimodular symmetric matrices,
\ie into the non-compact Riemannian symmetric space $SL(k, {\Bbb
R})/SO(k)$.  This space is known to have negative sectional curvature.

\subsubsection{A Bochner identity} 

   To proceed, we need to apply a ``Bochner Identity.'' To obtain it
we suppose, more generally, that a field $\phi^A(x): B \rightarrow N $
takes its values in some Riemannian target manifold $N$ with metric
$G_{AB}(\phi)$ and potential function $W(\phi)$.  In our case $\phi$
corresponds to the field of unimodular matrices ${\hat h}_{mn}$, and
thus $N$ is $SL(k,{\Bbb R})/SO(k)$ with its $SL(k,{\Bbb R}
)$-invariant metric, and $W(\phi)$ vanishes.  Actually we could
include the field $U$ and then $N= GL(k, {\Bbb R})/SO(k)= {\Bbb R}_+
\times SL(k, {\Bbb R})/SO(k)$ with the product metric.  In what
follows we retain $W(\phi)$ and assume that $\phi$ is coupled to the
metric $g_{\mu \nu}$ on $B$ in the standard fashion.

   The Bochner  identity tells us that:
\bea
&&(\ft12 G _ {AB}\, {\partial \phi^A \over \partial x^\alpha} \,
{\partial \phi^B \over \partial x^\beta} \, g^{\alpha
\beta} )^{;\mu}{}_{;\mu} = \phi^{A ; \alpha ; \beta} \, \phi^{B}
{}_{ ; \alpha ; \beta} \, G_{AB} \nn\\
&&+ \phi^{A;\alpha}\, ( G_{AB}\,  R_{\alpha \beta} - g_{\alpha \beta}
\, R_{ACBD} \, \phi^C_{; \mu} \, \phi^D_{; \nu} \, g^{\mu \nu} ) \,
\phi^{B ;\beta}\nn\\
&&+ \phi^{A ; \alpha} \, ( \phi^{B ;\beta}{}_{;\beta} )_{; \alpha}
\, G_{AB}\,,
\label{eqno51}
\eea
where all covariant derivatives are covariant with respect to the
spacetime metric $g_{\alpha \beta}$ and the target-space metric
$G_{AB}$, in the manner described  by \cite{Misner}.  The field
equations are:
\be
\phi^{A ; \beta}{}_{; \beta} = G^{AB} \, \nabla_B\,  W\,.
\label{eqno52}
\ee

   It is important to realise that in (\ref{eqno52}), the semi-colon
includes a contribution from the pull-back of the Levi-Civita
connection of $N$ to the spacetime manifold $B$ . More precisely,
$\partial_\mu \phi ^A$ is a section of the bundle: $T^* B \otimes
\phi^\star TN$ and $;$ denotes the connection on this bundle
associated to the Levi-Civita connections on $N$ and $B$.  On the
other hand, since it is acting on a scalar, the operator on the left
hand side of (\ref{eqno51}) is the usual Laplacian on $B$ with respect
to the metric $g_{\alpha \beta}$.

   The Einstein field equations read
\be
R_{\alpha \beta} = [ G_{AB}\,  \phi^A_{; \alpha}\,
\phi^B_{; \beta} + g_{\alpha \beta} \, W(\phi)]
\label{eqno53}
\ee
Now if we integrate (\ref{eqno51}) over $B$, dropping the boundary
term and assuming that $W(\phi)=0$, and if we assume that the Ricci
tensor of $B$, \ie $R_{\alpha \beta}$, is non-negative and that the
sectional curvatures of $N$ are non-positive, we see that the map
$\phi$ must be constant. This completes the proof of our splitting
theorem.

\subsection{Reduction on a unimodular Lie group}

    In this subsection turn briefly to case when the torus group $T^k$
is replaced by a general unimodular group $G$ of dimension $k$ with
left-invariant Cartan-Maurer forms $\lambda^n$ and structure constants
$C_l{}^m{}_n$.  To say the group is unimodular is to say that the
adjoint action preserves volume, or more concretely,
\be
C_l{}^m{}_m=0\,.
\ee
In the context of the Bianchi classification of three-dimensional Lie
algebras, the unimodular algebras are called Class A and the
non-unimodular algebras are called Class B.  It is known that
Kaluza-Klein reduction on a unimodular group is consistent in the
technical sense. It is useful to note that the contraction of a
unimodular group is itself unimodular.

   In the metric ansatz we replace $(dy^m+ A^m)$ by $(\lambda^m+
A^m)$, where now $A^m$ are $\frak{g}$-valued one forms.  The
Lagrangian must now be modified, since $F^m_{\mu \nu}$ is a
non-abelian curvature, and $\partial_\mu \, {\hat h}_{mn}$ must be
replaced by ${\cal D}_\mu\, {\hat h}_{mn}$ where ${\cal D}_\mu$ is the
gauge-covariant derivative acting on the symmetric tensor
representation ${\frak{g}} { \times^S } { \frak{g}}$. The quantity
$\partial_\mu U$ appears unchanged, but a potential term now arises,
of the form:
\be W=\ft14
 \exp({-2U \over d-2})\,
C^a{}_{bc}\,
\bigl ( 2 C^b{}_{ad } \,  h^{cd} + C^e{}_{fd}\,  h_{ad}\,  h^{bf}
\, h^{cd}\bigr ),
\ee
where $h_{mn} = \exp{2U \over k} {\hat h}_{mn}$.

   Note that both (\ref{Earnshaw}) and (\ref{current}) are modified,
and so one cannot immediately draw the same conclusions as before.
However, if the group $G$ admits a suitable circle subgroup, as it
does for the Heisenberg group or the Euclidean group, then more
can be said, as we shall show in the next subsection.

\subsection{Circle splitting; Heisenberg and Euclidean groups} 

   It is worth remarking to begin with that the issue of boundary
terms is a non-trivial one. Consider the case when the torus is a
circle, \ie $k=1, T^1=S^1=SO(2)=U(1)$, for which there is a single
Killing vector $K$. We might be tempted to use the identity for $U$
\ref{Earnshaw}, or use the covariant identity
\be \nabla^2 R^2 =|\nabla K | ^2, 
\ee 
with $R^2= |K |^2$, to attempt to prove that {\sl any} circle which
tends to constant length at infinity must split. But this is clearly
not always true. Indeed it is not true even if the Killing vector has
no fixed points and the length of the circle is bounded below by a
positive constant. In previous work examples of such phenomena have
already been exhibited (see, for example, \cite{munify}), in the case
of asymptotically locally conical (ALC) metrics of $G_2$ holonomy.
The circle in question could be identified with the M-theory circle,
and $R$ was related to the string coupling constant $g$ by $R\propto
g^{2 \over 3}$.  In these cases the boundary terms definitely do not
vanish. However, the examples of interest in the present paper
approach the flat metric at infinity at a fast enough rate that the
boundary terms do vanish, and hence any circle must flat.

   This means that we can extend our considerations to the case when
we contract $SU(2)$ not to the abelian group $T^3$, but to a
non-abelian group such as the Euclidean group $E(2)$ or the Heisenberg
group. In both cases there are circle subgroups, and we can apply our
results to any such circle subgroup, with the conclusion that they
must split. But if they do, it means that in fact the group we thought
was non-abelian is in fact abelian, \ie is $T^3$, and moreover the torus
must split. Thus the failure to find regular solutions with
$SU(2)\times T^3$ isometry group is part of a more general phenomenon
and not an artefact of performing too drastic a group contraction. In
a later section we shall exhibit some explicit metrics illustrating
what goes wrong.

\section{Special Lagrangian and  associative fibrations by tori}

     As an illustration of the above remarks, we can use some previous
results on cohomogeneity one metrics with special holonomy, in the
case of two-step nilpotent groups \cite{gilupost}.  All the metrics
are incomplete, and none is asymptotically Euclidean.  Nevertheless,
they provide instructive local models of the global problems with such
fibrations.  They may also provide local models of the situations
envisaged in \cite{guyaza}.

\subsection{Special Lagrangian tori in $SU(3)$ manifolds}

   Here we refer to section(4.1.2) and the metric (56) of
\cite{gilupost}:
\bea
ds_6^2 &=& H^2\, dy^2 + H^{-1}\, (dz_1 + m\, z_4\, dz_3)^2 + H^{-1}\,
(dz_2 + m\, z_5\, dz_3)^2 \nn\\
&&+ H^2\, dz_3^2 + H\, (dz_4^2 + dz_5^2)\,,\label{cy2met}
\eea
where $H$ is linear in $y$.  The Killing vector fields ${\partial
\over \partial z_i}$ with $i=1,2,3$ generate a torus action with
metric $ h_{mn}= {\rm diag} ( H^{-1}, H^{-1}, H^2)$. The torus has
constant volume and thus $U={\rm constant}$. The base $B$ has
coordinates $(y, z_4, z_5)$ and has topology ${\Bbb R} \times {\bf
R}^2$ if no identifications in $z_4$ and $z_5$ are made.

   The torus is clearly Lagrangian since by (59) of \cite{gilupost}
the symplectic form restricted to it vanishes. It is special because
in terms of the complex coordinates $\zeta_1\equiv z_1+\im\, H\, z_4$,
$\zeta_2\equiv z_2+\im\, H\, z_5$ and $\zeta_3=y+\im\, z_3$, the first
two are real and the last is pure imaginary along the torus, implying
that the holomorphic 3-form restricted to the torus has a constant
phase.

\subsection{Associative tori in $G_2$ manifolds}

   Here we refer to section(4.2.2) and the metric (75) of
\cite{gilupost}:
\bea
ds_7^2 &=& H^3\, dy^2 + H^{-1}\,(dz_1-m\, z_5\, dz_6)^2 + H^{-1}\, (dz_2
 +m\, z_4\, dz_6)^2 \nn\\
&&+ H^{-1}\, (dz_3 + m\, z_5\, dz_4)^2
+ H^2 (dz_4^2 + dz_5^2 + dz_6^2)\,.\label{7met2}
\eea
The Killing vector fields ${\partial \over \partial z_i}$ with
$i=1,2,6$ generate a torus action with metric $ h_{mn}= {\rm diag} (
H^{-1}, H^{-1}, H^2)$.  The torus has constant volume and thus $U={\rm
constant}$. The base $B$ has coordinates $(y, z_3, z_4, z_5)$ and has
topology ${\Bbb R} \times {\bf R}^3$ if no identifications in $z_3$,
$z_4$ and $z_5$ are made.

   The torus is associative since, by (76) of \cite{gilupost}, the
associative 3- form restricted to it gives its volume form. The base
space $B$ of the foliation is, by Hodge duality, a co-associative
4-fold.  As explained in \cite{gilupost}, these solutions are limiting
forms of complete non-singular solutions with isometry group $SU(2)
\times SU(2)$, and hence they are solutions of the equations obtained
by contraction of the general set of equations given earlier in this
paper.

\subsection{Associative fibrations of  $G_2$ manifolds constructed from
K3 surfaces}

   Another $G_2$ metric was given in section 4.2.1 of \cite{gilupost},
which was constructed from a nilpotent group which is a $T^2$ bundle
over $T^4$. It was also pointed out in section 6.2 of \cite{gilupost}
that it admits of an immediate generalisation in which the 4-torus is
replaced by a K3 surface. The metric is given locally by (126) of
\cite{gilupost} and is
\be 
ds^2 = H^4 dy ^2 + H^{-1} ( dx^1 + A^1 ) ^2 + H^{-1} ( dx^2 +
A ^2) ^2 + H^2 ds_4^2 , 
\ee 
where $ds_4^2$ is a K3 metric and $J^i= dA^i$, $i=0,1,2$ are the three
K\"ahler forms with vector potentials $A^i$ on the K3 surface, and $H$
is a harmonic function of $y$ which may be taken without loss of
generality as $H=y$. Globally we have a foliation by the sum of two
line bundles over the K3 surface whose curvatures are given by the two
K\"ahler forms $J^1$ and $J^2$. The global existence places some
restrictions on the K3 surface. An alternative description is to say
that we have a fibering over $K3$ by fibres $F$ with coordinates $(y,
x^1, x^2)$ which are ${\Bbb R}_+ \times T^2$.

   The associative 3-form $\psi_{(3)}$ was given in (128) of
\cite{gilupost}:
\be 
\psi_{(3)}= {\hat e}^0 \wedge
{\hat e} ^1 \wedge{ \hat e} ^2 + H^2 {\hat e} ^0 \wedge J^0 -{\hat
e}^1 \wedge J^2 + {\hat e ^2 } \wedge J^1.
\ee
Because the restriction of the associative 3-form to the fibres
gives its volume form, it follows that the fibres $F$ are
associative and hence by Hodge-duality that the $K^3$'s are
co-associative.


\begin{thebibliography}{99}      

\bm{brysal} R.L. Bryant and S. Salamon, {\it On the construction of
some complete metrics with exceptional holonomy}, Duke Math. J. {\bf
58}, 829 (1989).

\bm{gibpagpop} G.W. Gibbons, D.N. Page and C.N. Pope, {\it Einstein
metrics on $S^3$, $\R^3$ and $\R^4$ bundles}, Commun. Math. Phys.
{\bf 127}, 529 (1990).

\bm{cglpspin7} M.~Cveti\v c, G.~W.~Gibbons,
H.~L\"u and C.~N.~Pope, {\it New complete non-compact Spin(7)
manifolds,} hep-th/0103155; {\it New cohomogeneity one metrics with
Spin(7) holonomy,} math.DG/0105119.

\bm{cglp6fun} 
M. Cveti\v c, G.W. Gibbons, H. L\"u and C.N. Pope,
{\it Supersymmetry M3-branes and $G_2$ manifolds,}
Nucl.\ Phys.\ {\bf B620}, 3 (2002), hep-th/0106026.

\bm{bragugogo} A. Brandhuber, J. Gomis, S.S. Gubser and S. Gukov,
{\it Gauge theory at large N and new G(2) holonomy metrics,}
Nucl. Phys. {\bf B611}, 179 (2001), hep-th/0106034.

\bm{brand} A. Brandhuber, {\it $G_2$ holonomy spaces from invariant
three-forms,} hep-th/0112113.

\bm{munify}  M. Cveti\v c, G.W. Gibbons, H. L\"u and C.N. Pope,
{\it A $G_2$ Unification of the Deformed and Resolved Conifolds},
hep-th/0112138, to appear in Phys. Lett. {\bf B}. 

\bibitem{cglpmcon}
M. Cveti\v c, G.W. Gibbons, H. L\"u and C.N. Pope,
{\it M-theory conifolds,} Phys.\ Rev.\ Lett.\  {\bf 88}, 
121602 (2002), hep-th/0112098.

\bm{hitch} N. Hitchin, {\it Stable forms and special metrics}, 
math.DG/0107101.

\bm{guyaza} S. Gukov, S-T. Yau and E. Zaslow, {\it Duality and
fibrations on $G_2$ manifolds}, hep-th/0203217.

\bm{gilupost} G.W. Gibbons, H. L\"u, C.N. Pope and K.S. Stelle,
{\it Supersymmetric domain walls from metrics of special holonomy,}
Nucl.\ Phys.\ {\bf B623}, 3 (2002),
hep-th/0108191.

\bm{acharaya} B.~S.~Acharya, {\it On realising N = 1 super Yang-Mills
in M theory}, hep-th/0011089.

\bibitem{atmava} M.~Atiyah, J.~Maldacena and C.~Vafa,
{\it An M-theory flop as a large $n$ duality},  J. Math. Phys. {\bf 42},
3209 (2001), hep-th/0011256.

\bm{atiwit} M. Atiyah and E. Witten, {\it M-theory dynamics on a
manifold of $G_2$ holonomy}, hep-th/0107177.

\bibitem{kanyas1} H. Kanno and Y. Yasui,
{\it On Spin(7) holonomy metric based on $SU(3)/U(1)$},
hep-th/0108226.

\bibitem{cglpg2spin7}
M. Cveti\v c, G.W. Gibbons, H. L\"u and C.N. Pope,
{\it Cohomogeneity one manifolds of Spin(7) and $G_2$ holonomy,}
hep-th/0108245.

\bibitem{gukspa} S. Gukov and J. Sparks,
{\it M-theory on Spin(7) manifolds. I,} Nucl. Phys. {\bf B625} (2002)
3, hep-th/0109025.

\bibitem{cglporient} M. Cveti\v c, G.W. Gibbons, H. L\"u
and C.N. Pope, {\it Orientifolds and slumps in $G_2$ and Spin(7) metrics,}
hep-th/0111096.

\bibitem{kanyas2} H. Kanno and Y. Yasui,
{\it On Spin(7) holonomy metric based on SU(3)/U(1). II},
hep-th/0111198.

\bibitem{hersfe} R. Hernandez and K. Sfetsos,
{\it An eight-dimensional approach to $G_2$ manifolds}, hep-th/0202135.

\bm{gukov} S. Gukov, K Saraikin and A. Volovich, unpublished notes.

\bm{begipapo} V.A. Belinsky, G.W. Gibbons, D.N. Page and C.N. Pope,
{\it Asymptotically Euclidean Bianchi IX metrics in quantum gravity},
Phys. Lett. {\bf B76}, 433 (1978).

\bm{syza} A. Strominger, S.-T. Yau and E. Zaslow, {\it Mirror
Symmetry is T-Duality,} Nucl.\ Phys {\bf B479}, 243-259, (1996),
hep-th/9606040.

\bm{Misner} C. W. Misner, {\it Harmonic maps as models for physical
theories, } Phys.\ Rev.\ {\bf D18} 4510-4524 (1978).

\end{thebibliography}
\end{document}